\documentclass[11pt,a4paper]{article}
\pdfoutput=1
\usepackage{jcapmod}
\usepackage{booktabs}
\usepackage[english]{babel}
\usepackage{amsmath,amssymb,amsbsy,amstext, amsthm, simplewick}
\usepackage{graphicx}
\usepackage{float}
\usepackage{amsfonts}
\usepackage{amssymb}
\usepackage{braket}
\usepackage{upgreek}
 \usepackage{exscale,relsize}
 \usepackage[makeroom]{cancel}
\usepackage{soul}

\allowdisplaybreaks[1]

\usepackage[top=2.5cm, bottom=3cm, outer=2cm, inner=2cm]{geometry}

\newcommand{\beq}{\begin{equation}}
\newcommand{\eeq}{\end{equation}}
\newcommand{\beqa}{\begin{align}}
\newcommand{\eeqa}{\end{align}}
\newcommand{\calD}{\mathcal{D}}
\newcommand{\calL}{\mathcal{L}}
\newcommand{\calK}{\mathcal{K}}
\newcommand{\calO}{\mathcal{O}}

\newcommand{\calW}{\mathcal{W}}

\newcommand{\p}{\partial}

\newcommand{\Kahler}{K\"{a}hler }
\newcommand{\Mp}{M_{\scriptscriptstyle P\ell}}

\begin{document}

\begin{titlepage}

\setcounter{page}{1} \baselineskip=15.5pt \thispagestyle{empty}

\bigskip\

\vspace{1cm}
\begin{center}

{\fontsize{20}{24}\selectfont  \sffamily \bfseries  Building Supergravity Quintessence Model}

\end{center}

\vspace{0.2cm}
\begin{center}
{\fontsize{13}{30}\selectfont Chien-I Chiang$^{\bigstar\diamondsuit}$ and Hitoshi Murayama$^{\bigstar\diamondsuit\spadesuit}$} 
\end{center}

\begin{center}

\vskip 8pt
\textsl{$^\bigstar$Berkeley Center for Theoretical Physics and Department of Physics, University of California, Berkeley, CA 94720, USA} 

\vskip 6pt

\textsl{$^\diamondsuit$Physics Division, Lawrence Berkeley National Laboratory
Berkeley, California 94720} 

\vskip 6pt

\textsl{$^\spadesuit$Kavli Institute for the Physics and Mathematics of the Universe (WPI), \\
University of Tokyo Institutes for Advanced Study, University of Tokyo, Kashiwa, Japan}

\vskip 7pt

\end{center}

\vspace{1.2cm}
\hrule \vspace{0.3cm}
\noindent {\sffamily \bfseries Abstract} \\[0.1cm]
It was recently pointed out that the cosmological constant (even metastable one) belongs to the so-called ``swampland" and hence cannot be obtained as the low-energy limit of string theory that requires $|\nabla V| > c V$.  If true, the dark energy needs to be described by an evolving scalar field, {\it i.e.}\/, quintessence with $w>-1$ within supergravity.  However, the large hierarchy between the supersymmetry breaking scale and the energy scale of dark energy imposes a challenge on building quintessence models in supergravity as the quintessence field typically acquires a mass of order the gravitino mass. We investigate two approaches to circumvent this obstacle. One is imposing shift symmetry to the quintessence sector, and we demonstrate any quintessence potential can be embedded into supergravity and the fifth force constraint gives little limit on quintessence field displacement, leading to possible observational signature $w>-1$. The structure is stable against quantum corrections.  A particular example can address the cosmic coincidence problem. The other approach is sequestered supergravity, and the stability requirement strongly constrains the form of the \Kahler potential and superpotential, and the quintessence field displacement is typically much smaller than Planck mass. In addition, to satisfy the fifth force constraint, the quintessence field displacement is further restricted in the sequestered case, requiring $c \ll 1$.

\vskip 10pt
\hrule
\vskip 10pt

\vspace{0.6cm}
 \end{titlepage}

\noindent \rule{\textwidth}{0.5pt}

\tableofcontents

\vspace{0.5cm}

\noindent \rule{\textwidth}{0.5pt}

\section{Introduction}
Ever since its discovery \cite{Perlmutter:1998np,Riess:1998cb}, the current accelerating expansion of the universe has been one of the major puzzles of modern physics and its cause is often dubbed dark energy as its very nature is still an mystery.  The simplest solution may be adding a pure cosmological constant to the Einstein-Hilbert action and indeed the $\Lambda$CDM model has described our universe quite well \cite{Aghanim:2018eyx}. Nevertheless, the physical origin of cosmological constant has remained obscured and the na\"ive theoretical expectation is about 120 orders of magnitude larger than the observed value \cite{Weinberg:1988cp}. To explain the value of cosmological constant, one may appeal to anthropic arguments \cite{Weinberg:1987dv,Garriga:1999bf}, whose recent resurgence stems from the string theory landscape \cite{Dasgupta:1999ss,Bousso:2000xa,Giddings:2001yu,Kachru:2003aw,Susskind:2003kw}. To date, cosmological constant problem remains one of the most challenging problem in fundamental physics. 

Cosmological constant problem aside, over the years many alternatives have been proposed to account for the accelerating expansion. Among various proposals, there is a class of models where the dark energy is attributed to a canonical scalar field named quintessence \cite{Ratra:1987rm,Wetterich:1987fm,Zlatev:1998tr}. For a review see \cite{Tsujikawa:2013fta}. Some early models of this kind posses tracker behavior where the evolution of the field at late time is insensitive to initial conditions and hence make them rather attractive. Yet, as the observations have significantly improved for the past decades, now such models are under strong pressure from the observational constraints \cite{Linder:2015zxa}. But regardless the initial condition problem and/or cosmic coincidence problem (why the energy density of matter and dark energy are comparable at present time) can be solved or not, one basic question we wish to know is whether dark energy is purely a constant or if it is dynamical and evolves over time. Thanks to the advancement in many cosmological observations like eBOSS, SuMIRe (HSC and PFS on Subaru), DESI, Euclid, WFIRST and many others in the near future, we will have better sensitivity to see if the equation of state parameter $w$ of dark energy has any deviation from $-1$, which is the case if dark energy is not a pure cosmological constant. From this perspective, quintessence models are phenomenological tools that help us describe dark energy if it is dynamical with $w>-1$ and very often  the vacuum energy contribution is assumed to be zero due to other mechanism. Certainly, regardless dark energy is a pure cosmological constant or not, one still needs to answer if vacuum energy contributes to dark energy and if so, how large it should be. Yet, these are ambitious problems and very likely a full theory of quantum gravity is required to completely solve the cosmological constant problem.  On the other hand, recently a constraint on scalar field potential from quantum gravity was proposed in \cite{Obied:2018sgi} which suggests that the de Sitter vacuum may belongs to the ``swampland", where models cannot be UV completed with consistent theory of quantum gravity,  while the quintessence models are still safe \cite{Agrawal:2018own}. This gives another motivation to reexamine quintessence model-building. 

 Obviously, the string theory requires supersymmetry and hence its low-energy limit must be studied within the supergravity (SUGRA) theory.  Therefore, quintessence models must be formulated within SUGRA.
For example, see \cite{Brax:1999gp,Copeland:2000vh,Brax:2006kg,Brax:2006dc,Brax:2009kd} for some previous works along this line.  One particular point we would like to emphasize and is the focus of this paper is that when building quintessence model in supergravity, it is necessary to consider the effect of supersymmetry (SUSY) breaking on the quintessence sector because even if one successfully constructs a quintessence model alone, the SUSY breaking effect will spoil the flatness of the potential. In particular, the mass scale of quintessence is at the order of current Hubble parameter $H_0 \sim 10^{-33}$ eV. On the other hand, quite often quintessence will acquire a mass that has the same order as the gravitino mass $m_{3/2}$ which, for example, is about TeV in gravity mediation models, way much larger then the mass scale of quintessence. This steepens the quintessence potential, yielding the field settles at the minimum in early time and one cannot distinguish it from a pure cosmological constant. 

To be more concrete, let us consider a simple model where the hidden and quintessence sector are separated in the \Kahler potential with the canonical form, 
\beq
K = z^* z + Q^* Q,
\eeq
where $z$ and $Q$ are the chiral superfields of the hidden and quintessence sector respectively. This is a natural assumption in the sense that one would expect the interaction between the hidden and quintessence sector is as minimal as possible so there should be no cross terms in the \Kahler potential. Similarly, we assume the two sectors are separated in the superpotential as well,
\beq
W= W_0(z) + W_1(Q).
\eeq
Given the \Kahler potential and superpotential, the F-term scalar potential then reads
\beq
V_F = e^{K/\Mp^2}\left[D_i W K^{i\bar{j}}D_{\bar{j}}W^* - \frac{3}{\Mp^2} |W|^2 \right], \label{SUGRAFpotential}
\eeq
where $i$ and $j$ sum over the two sectors and
\beq
D_i W \equiv \frac{\p W}{\p \Phi^i} + \frac{W}{\Mp^2} \frac{\p K}{\p \Phi_i}.
\eeq
Here $\Mp$ is the reduced Planck mass $\Mp \equiv 1/\sqrt{8\pi G}$. Among various terms in the potential, there is a quadratic term of quintessence that couples to the superpotential of the hidden sector,
\beq
V \supset \frac{|W_0|^2}{\Mp^4} |\p_Q K|^2  = \frac{|W_0|^2}{\Mp^4} |Q|^2. \label{PolonyiMassTerm}
\eeq
As the gravitino mass $m_{3/2}$ is related to the superpotential  by $\braket{|W_0|^2} \sim m^2_{3/2} \Mp^4$, we see that
\beq
V \supset m^2_{3/2}|Q|^2.
\eeq
Due to the large hierarchy between the gravitino mass scale and the current Hubble scale, such term will make  quintessence roll down to the minimum and stick at there at a very early time, regardless how flat the potential is in the quintessence sector \textit{alone}. Observationally, quintessence then acts like a non-dynamical cosmological constant.

If we wish to construct a quintessence model that can be observationally distinguishable from a pure cosmological constant, for example having a time-varying equation of state in the present epoch, then one needs to prevent the quintessence sector from acquiring such gravitino mass. One known method is to impose shift symmetry to the quintessence sector \cite{Brax:2009kd}. We will review this in Sec.\ref{sec:ShiftQ}, emphasizing that one can incorporate quintessence with all kinds of potential into supergravity using shift symmetry. As a particular example, we will show that a hidden supersymmetric QCD axion \cite{ArkaniHamed:2000tc} can naturally play the role of quintessence and be embedded into SUGRA.  The cosmic coincidence problem is also ameliorated in such scenario. After reviewing the case with shift symmetry, in Sec.\ref{sec:SequesteredQ} we will show our attempt to construct a quintessence model where the quintessence and hidden sector are \textit{sequestered}, inspired by the brane-world scenario \cite{Randall:1998uk}. In such sequestered scenario, quintessence is protected from the SUSY breaking at least at the tree level, and it is possible to construct quintessence models of the small field type where the quintessence was frozen by Hubble damping for most of the time and only thawed recently. Yet, the constraint from the fifth force remains strong in this case and quintessence field value is limited in a tiny range, rendering it challenging to observationally distinguish such model from cosmological constant. However, in the phenomenological allowed range, exactly because of the small field displacement, the quantum correction beyond the tree level is well suppressed and the model is consistent from the effective field theoretic point of view. On the other hand, in the case with shift symmetry, the fifth force constraint is avoided. We conclude in Sec.\ref{sec:Discussion}.

\section{SUGRA Quintessence with Shift Symmetry}\label{sec:ShiftQ}
We first review the construction of quintessence model in SUGRA where a shift symmetry is imposed on the imaginary part of the quintessence sector. Particularly, the \Kahler potential has the form
\beq
K = z^* z + h(Q+Q^*). \label{ShiftKahler}
\eeq
where $h$ is an arbitrary function of $Q+Q^*$ with nonvanishing second derivative. 
We also make the two sectors separated in the superpotential,
\beq
W = W_0(z) + W_1(Q).
\eeq
The F-term potential then has the form
\beq
V_F = e^{\frac{K}{\Mp^2}} \left\{\left|\frac{\p W_0}{\p z} + \frac{1}{\Mp^2} z^*(W_0 + W_1) \right|^2 
+ \frac{1}{h''}\left| \frac{\p W_1}{\p Q} + \frac{1}{\Mp^2} h' (W_0 + W_1)  \right|^2
-\frac{3}{\Mp^2} |W_0+W_1|^2 \right\}
\eeq
where the prime on $h$ denotes the derivative with respect to its argument. Below we will denote the real and imaginary part of the quintessence sector as $r$ and $q$ respectively,
\beq
Q = r+ iq.
\eeq
Note that because of the large hierarchy between the SUSY breaking scale and the energy scale of dark energy, the dynamics of SUSY breaking will not be affected by the quintessence sector. To be more precise, we consider the superpotential of the quintessence sector of the form 
\beq
W_1(Q) = \Lambda^3 \, \calW_1\left(\frac{Q}{\Mp} \right), \label{W1}
\eeq
where $\Lambda$ is the energy scale of dark energy and $\calW_1$ is a holomorphic function of $Q/\Mp$. Note that the shift symmetry is broken by the superpotential which is inevitable as superpotential has to be holomorphic. This gives quantum corrections to the \Kahler potential that breaks shift symmetry. However, radiative stability is controlled by the smallness of $\Lambda$. We can also consider superpotentials that involve more parameters, as long as these parameters are smaller than $\Lambda$. For simplicity and minimality, we consider superpotentials of the form of Eq.(\ref{W1}).  The coupling between the hidden sector and $W_1$ will not affect the dynamics of hidden sector because of the smallness of $\Lambda$. The only interaction between the hidden and quintessence sector that does not involve $W_1(Q)$ has the form
\beq
\frac{1}{h''}\frac{|W_0|^2}{\Mp^4} h'^2 \sim m^2_{3/2} \frac{h'^2}{h''}. \label{quadraticterm}
\eeq
As $h$ only depends on the real part $r$ and the gravitino mass $m_{3/2}$ is much greater than the dark energy energy scale, this term sets the vacuum expectation value (vev) of $r$ such that $\braket{h'}=0$. On the other hand, the interaction terms between the hidden sector and quintessence sector that involve $W_1$ are suppressed by $\Lambda$. Hence, when determining the vev of the hidden sector, it is sufficient to only consider the terms depending on $z$ only, 
\beq
V_{\cancel{\text{SUSY}}} = e^{K/\Mp^2} \left\{\left|\frac{\p W_0}{\p z} \right|^2
+ \frac{1}{\Mp^2}\left( z^* \frac{\p W^*_0}{\p z^*} W_0  + z \frac{\p W_0}{\p z} W^*_0\right) + \frac{1}{\Mp^4} |z^2| |W_0|^2 - \frac{3}{\Mp^2}|W_0|^2 \right\}.
\eeq
Assuming $\braket{z}$ and $\braket{W_0}$ are real, the potential of the quintessence sector then has the form
\begin{align}
V_{\text{Quin}} & = e^{\frac{\braket{z}^2 + h}{\Mp^2} }  \Bigg\{ \frac{|\braket{W_0}|^2}{\Mp^4} \frac{h'^2}{h''}
+ \frac{1}{\Mp^2}\left(\left< z \frac{\p W_0}{\p z} \right> + \frac{\braket{|z|^2}}{\Mp^2} W_0 - 3 \braket{W_0} \right) \left(W_1 + W^*_1 \right)  \nonumber \\
       & \hspace{2cm}
       		+ \frac{1}{\Mp^2} \frac{h'}{h''} W_0  \left(\frac{\p W_1}{\p Q} + \frac{\p W^*_1}{\p Q^*} \right)
		+ \frac{W_0}{\Mp^4} \frac{h'^2}{h''} (W_1 + W^*_1)  \nonumber \\
       & \hspace{3cm}
       	+ \left| \frac{\p W_1}{\p Q} \right|^2
	+ \left(\frac{|\braket{z}|^2}{\Mp^4} + \frac{1}{h''} \frac{h'^2}{\Mp^4} - \frac{3}{\Mp^2} \right) |W_1|^2
	+ \frac{1}{\Mp^2} \frac{h'}{h''} \left(\frac{\p W_1}{\p Q} W_1^* + \frac{\p W^*_1}{\p Q^*} W_1\right) 
	 \Bigg\} \label{Eq102}
\end{align}
Despite many terms shown in the equation above, it can be largely simplified. In the first line, the first term is exactly Eq.(\ref{quadraticterm}) which has a energy scale much larger than the other terms as $\braket{W_0} \sim m_{3/2} \Mp^2$. This term sets the vev of $r$ such that $\braket{h'}=0$, and hence the second line can be dropped. For the terms in the third line, they are all at the order of $\calO(W^2_1)$, which has an energy scale of $\Lambda^6/\Mp^2$, where $\Lambda$ is defined in Eq.(\ref{W1}). As we wish the potential to be at the order of current dark energy scale, and the leading term is the second term in the first line, we have
\beq
m_{3/2} \Lambda^3 \sim \frac{M^2_{\cancel{\text{SUSY}}} \Lambda^3}{\Mp} \sim \Mp^2 H^2_0 \quad \Rightarrow 
\Lambda \sim \Mp \left(\frac{H^2_0}{M^2_{\cancel{\text{SUSY}}}} \right)^{1/3}. \label{gravitinomassLambda}
\eeq
where $M_{\cancel{\text{SUSY}}}$ is the SUSY breaking scale. The large hierarchy between $\Lambda$ and $\Mp$ is a manifestation of cosmological constant problem. Because of this large hierarchy, the third line in Eq.(\ref{Eq102}) is largely suppressed. For example, in the case of gravity-mediation scenarios, $M^2_{\cancel{\text{SUSY}}}\sim 10^{21}(\text{GeV})^2$,  hence $\Lambda \sim 10^{-35}\Mp$ and the third line in Eq.(\ref{Eq102}) has a energy scale of $10^{-210}\Mp^4$, which is 90 orders of magnitude smaller than that of the second line. In low-scale SUSY breaking models, $M^2_{\cancel{\text{SUSY}}}\sim (\text{TeV})^2 \sim 10^{-30}\Mp^2$, yielding $\Lambda \sim 10^{-30}\Mp$. The third line in Eq.(\ref{Eq102}) therefore has a energy scale of $10^{-180}\Mp^4$, which is 60 orders of magnitude smaller than that of the second line. In conclusion, we can drop terms in the third line in either cases and the potential of the quintessence sector has the following simple form when shift symmetry is imposed on the \Kahler potential of the quintessence sector,  
\beq
V_{\text{Quin}} =  e^{\frac{\braket{z}^2 + h}{\Mp^2}}   \frac{1}{\Mp^2}\left(\left< z \frac{\p W_0}{\p z} \right> + \frac{\braket{|z|^2}}{\Mp^2} W_0 - 3 \braket{W_0} \right) \left(W_1 + W^*_1 \right). \label{shiftV}
\eeq
For example, suppose the hidden sector have a superpotential of Polonyi type,
\beq
W_0 =  \mu \Mp (z+ \beta).
\eeq
where $\mu$ is a parameter of mass dimension one. Requiring the hidden sector contribute zero vacuum energy $V_{\cancel{\text{SUSY}}}(\braket{z})=0$, one finds
\beq
\braket{z} = (\sqrt{3}-1) \Mp \quad\text{and} \quad \beta= (2-\sqrt{3})\Mp.
\eeq 
The gravitino  mass is given by
\beq
m_{3/2}=e^{2-\sqrt{3}} \mu.
\eeq
Assuming the quintessence sector has the canonical \Kahler potential with shift symmetry,
\beq
h=\frac{1}{2} \big(Q+ Q^* \big)^2,
\eeq
the potential of the quintessence then has a very simple form
\beq
V_{\text{Quin}} = - \sqrt{3} \, e^{2-\sqrt{3}}m_{3/2} (W_1 + W^*_1). \label{shiftVcanonical}
\eeq
We see that for any quintessence model with a potential $V_{\text{Quin}}$, one can embed it into supergravity by making the real part of the superpotential proportional to the potential.

Note that the shift symmetry of the \Kahler potential will be inevitably broken by superpotential, as superpotentials are holomorphic. We can estimate the effect of this shift symmetry breaking by considering the quantum correction to the \Kahler potential from superpotential coupling. In particular, considering the leading order correction, the cubic interaction term in the superpotential $W_1\supset (\Lambda^3/\Mp^3) Q^3$ yields a loop correction to the \Kahler potential $K_{\rm q.c.}$ of the form
\beq
K_{q.c.} \sim \frac{1}{16\pi^2} \frac{\Lambda^6}{\Mp^6} |Q|^2. \label{shiftasymcorrec}
\eeq
Even though such correction breaks the shift symmetry in the \Kahler potential, because of the small coupling in the superpotential, i.e. the large hierarchy between $\Lambda$ and $\Mp$, this does not spoil the flatness of the quintessence potential. Indeed, as the scalar potential has the form $e^{K/\Mp^2} [\cdots ]$ as shown in Eq.(\ref{SUGRAFpotential}), the shift asymmetric correction Eq.(\ref{shiftasymcorrec}) leads to a mass term
\beq
V \supset \frac{1}{16\pi^2}\frac{\Lambda^6}{\Mp^6} \frac{V_{\rm Quin}}{\Mp^2} |Q|^2 \sim H^2_0 \left(\frac{\Lambda^6}{\Mp^6}\right)  |Q|^2, \label{HubbleInducedMass}
\eeq
where the mass $(\Lambda^3/\Mp^3) H_0$ is much smaller than the current Hubble scale $H_0$ and is therefore harmless. One may also worry the coupling to the SUSY breaking sector like Eq.(\ref{PolonyiMassTerm}) which yields
\beq
V \supset \frac{|W_0|^2}{\Mp^4} |\p_Q K|^2 \sim m^2_{3/2} \left(\frac{\Lambda^6}{\Mp^6} \right)^2 |Q|^2
\sim H^2_0 \left(\frac{H^2_0}{\Mp^2} \right)\left(\frac{\Lambda^6}{\Mp^6} \right) |Q|^2,
\eeq
where we have used $m_{3/2} \Lambda^3 \sim H^2_0 \Mp^2$ as given in Eq.(\ref{gravitinomassLambda}). We see that this contribution is even smaller than that in Eq.(\ref{HubbleInducedMass}) with an additional suppression $H^2_0/\Mp^2$. Overall we see that even though the shift symmetry of the \Kahler potential will be radiatively broken by the superpotential, the effect is negligible and the potential Eq.(\ref{shiftVcanonical}) is protected from quantum corrections.

A quintessence sector with shift symmetry and the right energy scale can naturally arise \cite{ArkaniHamed:2000tc}, based on the observation that the energy scale of dark energy is related to the electroweak scale $M_{EW}$ and Planck scale by
\beq
\rho^{1/4}_{DE} \sim \frac{M^2_{EW}}{\Mp}.
\eeq  
In particular, assume SUSY is broken at the TeV scale by an order parameter chiral superfield $\langle S \rangle = \theta^2 M^2_{EW}$ and there is a hidden supersymmetric QCD (SQCD) sector $\Psi$ with $SU(N_c)$ gauge group and $N_f$ flavors that couples to SUSY breaking sector and the observable sector only through Planck-suppressed interactions. Once the SUSY is broken, the hidden sector quarks acquire a mass through the operator
\beq
\int d^4 \theta \frac{S^*}{\Mp} \tilde{\Psi} \Psi
\eeq
and hence the masses of the hidden quarks are of the order of $m_{\Psi}\sim M^2_{EW}/\Mp$. Similarly, the hidden gluino acquires a mass of the order of $m_\lambda /g^2 \sim M^2_{EW}/\Mp$ through the operator 
\beq
\int d^2 \theta \frac{S}{\Mp} \calW_\alpha \calW^\alpha = \frac{M^2_{EW}}{\Mp} \lambda \lambda.
\eeq 
Assuming the gluino mass is somewhat smaller than the others and $3N_c - N_f \ll N_f$, then the strongly coupled scale of the hidden SQCD is about the same as the hidden quark mass, 
\beq
\Lambda \sim m_{\Psi} \sim  \frac{M^2_{EW}}{\Mp},
\eeq
because the sector becomes strongly coupled quickly after the hidden quarks decoupled. With the gluino condensation, the axion $Q$ associated with the SQCD then has a superpotential 
\beq
W_{\text{axion}}  = \Lambda^3 e^{-Q/\Mp}.
\eeq
Plugging this back to Eq.(\ref{shiftVcanonical}) and an appropriate tuning of the cosmological constant contribution yield the usual cosine-type potential
\beq
V = \left(\frac{M^2_{EW}}{\Mp} \right)^4 \left[1 - \cos\left(\frac{q}{\Mp}\right) \right].
\eeq

The above paradigm is intriguing as it naturally connects the dark energy energy scale with the other two important physical scales, $M_{EW}$ and $\Mp$, and the cosmic coincidence problem can also be explained \cite{ArkaniHamed:2000tc}. In addition, as the quintessence sector is essentially the hidden SQCD axion, we have the shift symmetry to protect quintessence from acquiring large gravitino mass when we embed it into SUGRA. Yet, one should notice that such paradigm requires the SUSY breaking scale to be exactly at the TeV scale, which means the gravitino mass is at the order of meV. An explicit construction of such kind of model is challenging and yet to be done.

\section{Quintessence in Sequestered Supergravity}\label{sec:SequesteredQ}
In this section we show another attempt of constructing quintessence model in supergravity. In particular, we consider the case where the hidden and quintessence sectors are \textit{sequestered} \cite{Randall:1998uk}, yielding a \Kahler potential of the form
\beq
K = -3\Mp^2 \ln\left( 1 - \frac{f(z, z^*)}{3\Mp^2}  - \frac{g(Q,Q^*)}{3\Mp^2} \right),
\eeq
where $f$ and $g$ are real functions of $z$ and $Q$ respectively. This form of \Kahler potential can be  originated from higher dimensional theory where the two sectors live on two separate 3-branes. For the superpotential, we have the same form as before, 
\beq
W = W_0(z) + W_1(Q).
\eeq
Like in the case of shift symmetry, we first work out potential of the hidden sector, which will tell us how SUSY breaking affects quintessence. Working out the scalar potential, we have
\begin{align}
V_F & =\frac{\calK^{-2}}{\calD}\left\{ g_{Q Q^*}\left[\calK_0  \left|\frac{\p W_0}{\p z} \right|^2 + \frac{1}{\Mp^2}\left(f_z \frac{\p W^*_0}{\p z^*} W_0 + c.c. \right) - \frac{3}{\Mp^2}f_{z z^*} |W_0|^2 \right]  \right. \nonumber \\
& \quad \hspace{8cm} \left. + \frac{\left(|g_Q|^2-g\,g_{Q Q^*}\right)}{3\Mp^2} \left|\frac{\p W_0}{\p z} \right|^2 + \calO(W_1) \right\} \label{VFhid}
\end{align}
where 
\beq
\calK\equiv 1- \frac{f}{3\Mp^2} -\frac{g}{3\Mp^2}, \quad \quad
\calK_0 \equiv 1-\frac{f}{3\Mp^2}, \quad \quad
\calD = \calK f_{z z^*} g_{Q Q^*} + \frac{f_{z z^*}|g_Q|^2}{3\Mp^2} + \frac{g_{Q Q^*}|f_z|^2}{3\Mp^2}. \label{Eq3.4}
\eeq
The subscripts of $f$ and $g$ denotes derivative with respect to the respective variables. Because of the large hierarchy between the energy scale of dark energy and SUSY breaking scales, terms proportional to $W_1$ are negligible when considering the dynamics of SUSY breaking. From Eq.(\ref{VFhid}) we can see that \textit{if} the condition
\beq
\left(|g_Q|^2-g\,g_{Q Q^*}\right) =0 \label{gcondition}
\eeq
is satisfied and the square bracket in the first line vanishes when $z$ lies at its minimum such that the hidden sector contributes zero vacuum energy, then the quintessence sector will not acquire a mass of SUSY breaking scale. The condition Eq.(\ref{gcondition}) means that $g(Q, Q^*)$ has to be in the canonical form
\beq
g(Q, Q^*) = QQ^*. \label{gQQ}
\eeq
Indeed, if we expand $g$ to higher orders in the form of
\beq
g(Q,Q^*) = |Q|^2 \left(1+ \alpha_1 \frac{|Q|^2}{\Mp^2} + \alpha_2 \frac{|Q|^4}{\Mp^4} + \cdots \right), \label{gQQ}
\eeq
where $\alpha_i$'s are dimensionless coefficients, then the first term of the second line in Eq.(\ref{VFhid}) will then be
\beq
  - \frac{|Q|^2}{3\Mp^2} 
 \left[\alpha_1 \frac{|Q|^2}{\Mp^2} + 4\alpha_2 \frac{|Q|^4}{\Mp^4}  \cdots \right] \left|\frac{\p W_0}{\p z} \right|^2
 \sim - \frac{1}{3} \left[\alpha_1 \frac{|Q|^2}{\Mp^2} + 4\alpha_2 \frac{|Q|^4}{\Mp^4}  \cdots \right] m^2_{3/2} |Q|^2.
 \label{QGcorrection}
\eeq
Unless $|Q|^2/\Mp^2 \ll 1$, we see that the quintessence sector acquires an effective quadratic term at the order of gravitino mass squared. In general, $g(Q, Q^*)$ need not be in the canonical form as \Kahler potential would be renormalized when quantum effect is taken into account. Even if the two sectors live on separate branes, gravity still mediates between the two and quantum gravity effect generically spoils the tree level \Kahler potential. Nonetheless, this can be regarded as  another manifestation of the cosmological constant problem in the sense that quantum effect naively yields an energy scale much larger than the dark energy energy scale. In fact, when we make the hidden sector contributes zero vacuum energy, it is also controlled only at the tree level. As tackling quantum gravity effect fully remains challenging, we choose to proceed with the assumption that the canonical form of $g(Q, Q^*)$ is preserved by some unknown mechanism.

Moving ahead, let us workout the potential of the hidden sector more explicitly. For the superpotential we adopt the Polonyi type as we did in the shift symmetry case, 
\beq
W_0 = \mu \Mp (z+\beta),
\eeq
where $\mu$ is a mass dimension one parameter of the SUSY breaking scale. 
For the \Kahler potential, we consider the form
\beq
f= |z|^2 + \lambda \frac{|z|^4}{3\Mp^2}.
\eeq
The quartic term is included because with the quadratic term alone the potential will not be bounded from below. In fact, to make the potential bounded from below, $\lambda$ has to be negative and we choose $\lambda = -1/4$ for convenience. Demanding the hidden sector contributes zero vacuum energy, one finds
\beq
\beta =  \frac{\left(3-\sqrt{5} \right)^{3/2}}{2\sqrt{6}}\,\Mp
\eeq
and the hidden sector field lies at the vev
\beq
\braket{z}=\sqrt{\frac{6}{5}\left( 3-\sqrt{5} \right)}\,\Mp.
\eeq

With the hidden sector settles at its vev and taking $g(Q, Q^*)= |Q|^2$, the potential for the quintessence sector then reads
\begin{align}
V_{\rm Quin} & =  \frac{1}{\braket{\calD} \left( \braket{\calK_0} - \frac{|Q|^2}{3\Mp^2}\right)^2}
\Bigg\{ \left<\calK_0 f_{zz^*} + \frac{|f_z|^2}{3\Mp^2} \right>  |W_{1Q}|^2 \nonumber \\
& \hspace{150pt} + \frac{\braket{f_{zz^*}}}{3\Mp^2} \Big( 3W_1 Q^* W^*_{1Q} + \text{c.c.} - |Q|^2 |W_{1Q}|^2 - 9 |W_1|^2\Big) \nonumber \\
& \hspace{200pt} \frac{1}{3\Mp^2} \left[\left< 3 f_{z z^*} W_0 - f_z W_{oz}^* \right> \left( Q W_{1Q} - 3 W_1 + \text{c.c.} \right) \right] \Bigg\}. \label{Qpotential}
\end{align}
Note that the potential approaches infinity when $Q \to \pm \sqrt{3\braket{K_0}}\Mp \simeq 1.47 \Mp$, and hence the field displacement of the quintessence field is confined within this range, rendering large-field type model impossible in the sequestering setup.  Even though such singularity can be apparently removed when we make the field redefinition so that the kinetic term in the Lagrangian becomes canonical, the prefactor outside the curly brackets still dominates and one can check that the slow-roll parameters are larger than unity in the large field region. In fact, since part of the prefactor is originated from the $e^K$ prefactor of the scalar potential, this is similar to the $\eta$-problem in inflationary model building in supergravity. We therefore focus on small-field type potential, where, for instance, the quintessence field rolls on a plateau. These types of models are generally sensitive to initial conditions, unlike tracker-type quintessence models where the field evolutions with wide range of initial conditions converge to a common trajectory. Despite this shortcoming, our goal is to investigate the possibility if quintessence models can be built in sequestered supergravity that lead to observational signature distinguishable from cosmological constant, hence we will bear with this initial condition problem.

The first thing we would like to ensure is that the potential is bounded from below. In Eq.(\ref{Qpotential}) the first and second line are at the order of $\calO(W^2_1/\Mp^2)$, while the third line is at the order of $\calO(\mu^2 W_1/\Mp)$. Because of the SUSY breaking scale $\mu$, the third line generically has a much larger energy scale and hence dominates the potential. However, the third line of Eq.(\ref{Qpotential}) is not positive-definite and there is always a direction in the complex field space where the potential approaches negative infinity when $|Q| \to  \sqrt{3\braket{K_0}}\Mp$.  It seems that we need multiple parameters in the superpotential with a large hierarchy among them to make the potential bounded from below.

Consider a superpotential for the quintessence sector of the form
\beq
W_1(Q) = \Lambda^3  \left(\frac{Q}{\Mp} \right)^n,
\eeq
which yields
\begin{align}
V_{\rm Quin} & =  \frac{1}{\braket{\calD} \left( \braket{\calK_0} - \frac{|Q|^2}{3\Mp^2}\right)^2}
\Bigg\{ - \frac{\braket{f_{zz^*}}}{3} (n-3)^2  \frac{\Lambda^6}{\Mp^2} \left(\frac{|Q|^2}{\Mp^2} \right)^{n}  \nonumber \\
& \hspace{130pt} + \left<\calK_0 f_{zz^*} + \frac{|f_z|^2}{3\Mp^2} \right> n^2 \frac{\Lambda^6}{\Mp^2} \left(\frac{|Q|^2}{\Mp^2} \right)^{n-1}\nonumber \\
& \hspace{150pt} \frac{\Lambda^3}{3\Mp^2} \left< 3 f_{z z^*} W_0 - f_z W_{oz}^* \right> (n-3)  \left[ \left(\frac{Q}{\Mp} \right)^n + \left(\frac{Q^*}{\Mp} \right)^n \right] \Bigg\}.
\end{align}
We see that the potential is positive-definite only when $n=3$. Hence when we consider superpotential of polynomial form, the highest order must be truncated at $n=3$. Going beyond cubic order will yield potential that is unbounded from below.  We therefore consider
\beq
W_1(Q) = \tilde{a} \left(\frac{Q}{\Mp} \right)^3 + \tilde{b} \left(\frac{Q}{\Mp} \right)^2 + \tilde{c} \left(\frac{Q}{\Mp} \right) + \tilde{d},
\eeq
which gives the potential of the form
\begin{align}
V_{\rm Quin} & = \frac{1}{\braket{\calD} \left(\braket{\calK_0} - \frac{|Q|^2}{3\Mp^2} \right)^2}
\Bigg\{9\left<\calK_0 f_{zz^*} + \frac{|f_z|^2}{3\Mp^2}\right>\frac{|\tilde{a}|^2}{\Mp^2} \frac{|Q|^4}{\Mp^4}  \nonumber \\
&  \hspace{130pt} -  \frac{1}{3\Mp^2} \left< 3 f_{z z^*} W_0 - f_z W_{oz}^* \right> \left[ \tilde{b}  \left(\frac{Q}{\Mp}\right)^2 + 2\tilde{c} \left(\frac{Q}{\Mp} \right)+3\tilde{d}  + \text{c.c.} \right]  \Bigg\}
\end{align}
Note that in order to make all the terms to have the same energy scale and describe dark energy, we need $\tilde{a}\sim 10^{-60} \Mp^3$ and $\tilde{b} \sim \tilde{c} \sim \tilde{d} \sim 10^{-105} \Mp^3$. There is a large hierarchy between $\tilde{a}$ and the other three parameters because of the coupling of the latter three with the SUSY breaking scale $\mu$, where $\mu^2/\Mp \sim \text{TeV}$. Also note that because of this large hierarchy, we have neglected the cross terms in the potential like $\tilde{a}\tilde{b}$, $\tilde{b}\tilde{c}$, etc. Define the real and imaginary part of $Q/\Mp$,
\beq
r\equiv \mathfrak{Re}\frac{Q}{\Mp}, \quad \quad s\equiv \mathfrak{Im}\frac{Q}{\Mp},
\eeq
and assume the parameters are real, the potential has the form
\begin{align}
V_{\rm Quin} & = \frac{1}{\braket{\calD} \left(\braket{\calK_0} - \frac{r^2+s^2}{3\Mp^2} \right)^2}
\Bigg\{9\left<\calK_0 f_{zz^*} + \frac{|f_z|^2}{3\Mp^2}\right>\frac{|\tilde{a}|^2}{\Mp^2} (r^2 + s^2)^2 \nonumber \\
&  \hspace{130pt} +  \frac{2}{3\Mp^2} \left< 3 f_{z z^*} W_0 - f_z W_{oz}^* \right> \left[ \tilde{b}\, s^2 - \tilde{b} \,r^2-2\tilde{c}\, r -3\tilde{d} \right]  \Bigg\}.
\end{align}
Notice that the imaginary part has even power and global minimum at $s=0$. Hence we can assume the imaginary part lies at its minimum and focus on the potential for the real part only.

Lastly, note that in supergravity the kinetic term of the scalar field is given by
\beq
\calL_{kin} =-K_{QQ^*}\p_\mu Q \p^\mu Q^* =  - \frac{\calK_0}{\left[\calK_0 - \frac{|Q|^2}{3\Mp^2} \right]^2} \p_\mu Q \p^\mu Q^*
\eeq
Because the imaginary part lies at the $s=0$, from now on we will take $Q$ as real. The kinetic term will be canonical after we make the field redefinition
\beq
\tilde{Q}=\sqrt{3} \Mp \tanh^{-1}\left[\frac{Q}{\sqrt{3\calK_0}\Mp} \right].
\eeq
In terms of the canonically normalized field, we arrive at the general potential of the quintessence field in the setup of sequestered supergravity:
\beq
V_{\rm Quin} = \rho_{\scriptscriptstyle DE} \cosh^4\left(\frac{\tilde{Q}}{\sqrt{3} \Mp } \right)
\left\{a^2 \tanh^4\left(\frac{\tilde{Q}}{\sqrt{3} \Mp } \right) - b \tanh^2\left(\frac{\tilde{Q}}{\sqrt{3} \Mp } \right)
+ c \tanh\left(\frac{\tilde{Q}}{\sqrt{3} \Mp } \right) + d \right\} \label{GeneralQV}
\eeq 
Here $a$, $b$, $c$, and $d$ are dimensionless parameters where $b>0$ to ensure the imaginary part of the field lies at zero. 

\begin{figure}[H]
\begin{center}
\includegraphics[width=\textwidth]{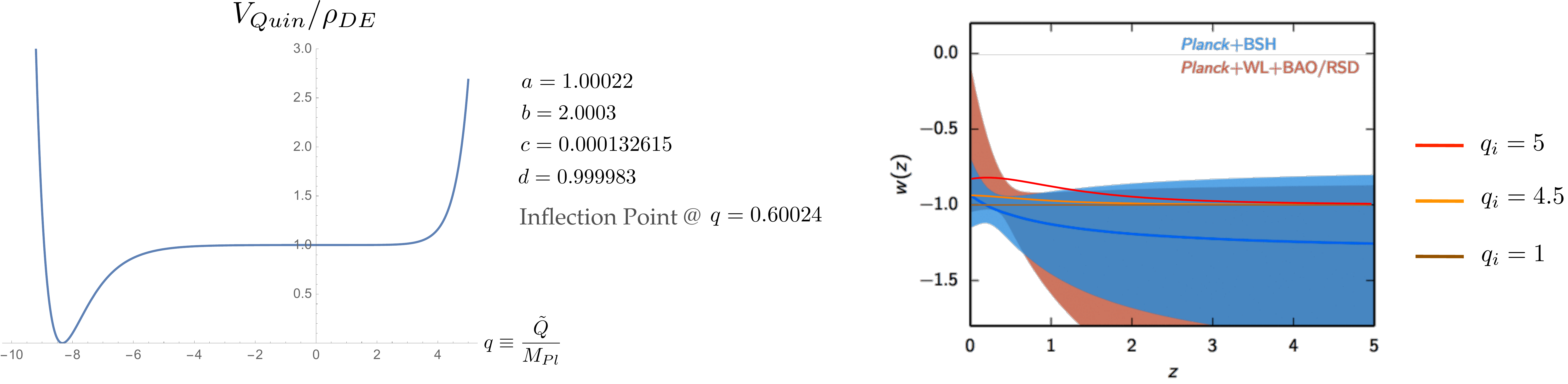}
\caption{(Left): Plot of the quintessence potential Eq.(\ref{GeneralQV}) with the given parameters. The potential has an inflection point at $q=0.60024$. The parameters are chosen in a way such that at the global minimum has $V=0$, while the quintessence field contributes the right amount of energy when it slow rolls on the plateau. (Right): The evolution of the equation of state parameter $w$ for the potential shown in the left panel with various initial conditions, in comparison with the $w(z)$ given in \cite{Ade:2015rim} which was reconstructed from observational data. We see that, for instance, with the quintessence field starting at $q_i=4.5$, one can have a interesting deviation from $w=-1$ that still satisfies current constraints. }
\label{fig:InflectionPoint}
\end{center}
\end{figure}

One possible scenario from this general form is a potential with long plateau on which the quintessence field slow-rolls. To obtain such kind of potential requires some fine-tuning of the parameters is required. In the left of Figure \ref{fig:InflectionPoint} we show an explicit example of such kind.

With a given potential of a long plateau, one also has to choose the initial condition for the field. The evolution of the quintessence field at late time, say redshift $z<5$, is insensitive to the initial velocity of the field. This is because no matter how large the initial velocity is, it would soon be damped away by the Hubble friction and remain frozen near at its initial position until the matter and radiation energy density become low enough. After the field started to roll, its equation of state parameter $w$ would gradually increase from $w=-1$. If the field started with an initial position $q_i$ on the steep slope, say $q_i=5$ for the potential shown in Figure \ref{fig:InflectionPoint}, $w$ may increase too much and exceed the observational bound. On the other hand, if the field started on the flat plateau, say $q_i=1$, $w$ would not deviate from $w=-1$ too much and act like a cosmological constant. The phenomenologically most interesting case happens when the field has an initial position at the junction of the flat plateau and the steep slope, say $q_i=4.5$.  The time evolution of the equation of state parameter with these three different initial conditions is shown on the right in Figure \ref{fig:InflectionPoint}.

Although considering the quintessence and SUSY breaking sector alone can lead to interesting observational signatures in the context of sequestered supergravity, things would unfortunately change when we consider the coupling between the quintessence and matter. Specifically, the constraint from the fifth force would require the quintessence field to have a nearly zero field value, resulting a equation of state parameter with little deviation from $w=-1$ and hence cannot be observationally distinguishable from a pure cosmological constant. To elaborate more on this, note that the fermion mass has the form
\beq
m_{u,d} = y\, e^{K/2\Mp^2} \, v_{u,d}, \label{fermionmass}
\eeq
where $y$ is the Yukawa coupling, $m_{u,d}$ are the mass of the $u$-type and $d$-type particles, and $v_{u,d}$ are the vev of $u$-type and $d$-type Higgs field $H_u$ and $H_d$. With the $\tilde{Q}$-dependence in the \Kahler potential, we effectively have a coupling between the quintessence and the matter sector, whose strength is determined by
\beq
\alpha\equiv \Mp \frac{\p \ln m}{\p \tilde{Q}}. \label{alpha}
\eeq
This interaction is often dubbed ``the fifth force". In sequestered scenario, we have $K = -3 \Mp^2 \ln \calK$, where $\calK$ is defined in Eq.(\ref{Eq3.4}). Hence the strength of the fifth force is given by
\beq
\alpha = -\frac{3}{2} \Mp \frac{dQ}{d\tilde{Q}} \frac{d\ln \calK}{dQ} = \sqrt{3} \tanh \left(\frac{\tilde{Q}}{\sqrt{3} \Mp} \right). 
\label{alphaSequester}
\eeq
Observational constraints on $\alpha$ from radar time-delay effect give stringent bounds on the strength of the fifth force. For instance, measurements made by Cassini spacecraft yield a bound of $\alpha^2 \lesssim 10^{-5}$ \cite{bertotti2003test}. By Eq.(\ref{alphaSequester}), this translates to the bound on the quintessence field
\beq
\left( \frac{\tilde{Q}}{\Mp} \right)^2 \lesssim 10^{-5}. \label{FifthForceConstraint}
\eeq
Such a tiny small range means the the equation of state parameter $w$ cannot be largely deviated  from $-1$. To see this, note that the current field velocity is related to the field displacement by
\beq
\dot{\tilde{Q}} \sim \Delta\tilde{Q} H_0.
\eeq
This is of course a rough approximation since the field doesn't roll for the whole time of the age of the universe. Yet the the difference should only be an order of one tenth. The equation of state parameter $w$ at late time can therefore be approximated as
\beq
w = \frac{\frac{1}{2} \dot{\tilde{Q}}^2 - V}{\frac{1}{2} \dot{\tilde{Q}}^2 + V} \sim 
\frac{\frac{1}{2} \Delta\tilde{Q}^2 H_0^2 - 3 \Mp^2 H^2_0}{\frac{1}{2} \Delta\tilde{Q}^2 H_0^2 + 3 \Mp^2 H^2_0}
\sim -1 + \frac{\Delta\tilde{Q}^2}{3\Mp^2}. \label{wDeviation}
\eeq
Hence, combining with the fifth force constraint, the deviation of equation of state parameter from -1 in sequestered supergravity is less than $10^{-5}$, making it challenging to be detected in current and near future observations.

Recently, a swampland conjecture regarding the shape of the scalar potential in any consistent theory of quantum gravity was put forward in \cite{Obied:2018sgi}, where the authors suggested that the potential of the scalar fields $\varphi_i$'s should satisfy the criterion 
\beq
\Mp |\nabla V| > c V, \label{dSConjecture}
\eeq
where $\nabla V$ is the gradient with respect to the scalar fields $\varphi_i$'s, and $c$ is a number of order $\calO(1)$. Due the fifth force constraint, the field value of $\tilde{Q}$ is confined to be closed to the origin and hence the potential Eq.(\ref{GeneralQV}) can be approximated as a linear potential of the form
\beq
V_{\rm Quin} = \rho_{\scriptscriptstyle DE}\left(1  + c\, \frac{\tilde{Q}}{\Mp}\right) .
\eeq
With this simple form, one can solve the evolution of $\tilde{Q}$ and the displacement $\Delta \tilde{Q}$ is given by
\beq
\Delta \tilde{Q} \sim c\, \Mp.
\eeq
This means that in order to satisfy the fifth force constraint Eq.(\ref{FifthForceConstraint}), we not only need to have an initial condition $\tilde{Q}_i^2 < 10^{-5}\Mp^2$, the slope of the potential also needs to satisfy $c^2 \lesssim 10^{-5}$. This will violate the swampland conjecture Eq.(\ref{dSConjecture}) if the number $c$ in the conjecture is of order one, and the sequestered scenario will be theoretically ruled out. Nonetheless, as pointed out in \cite{Dias:2018ngv}, the swampland conjectures should be regarded \textit{parametrically} and the number $c$ does not have to be of order one. Indeed, using effective field theoretic arguments, the authors in \cite{Dias:2018ngv} argued that $c$ should be the order of $m_\ell/ m_h$, where $m_\ell$ is the mass of light particle considered, and $m_h$ is the mass of the lightest heavy particle that one integrates out. Therefore, if the hierarchy between $m_\ell$ and $m_h$ is large enough, namely $(m_\ell/ m_h)^2 \lesssim 10^{-5}$, the sequestered scenario can still satisfy the swampland conjecture.

\section{Discussion and Conclusion}\label{sec:Discussion}
In this note we have discussed quintessence model building in supergravity. We stressed that there are two main issues when trying to construct supergravity quintessence models that are observationally indistinguishable from a pure cosmological constant. Firstly, for any realistic models, it is necessary to consider the effect of SUSY breaking which often gives quintessence a mass at the scale of the gravitino mass, which is much larger than that of the current Hubble scale. This renders the potential too steep such that quintessence settles at the minimum in the very early time and acts like a pure cosmological constant. One can avoid this problem by imposing shift symmetry on the quintessence sector, and an advantage of this approach is that it is much easier to embed any quintessence potential in this framework -- the quintessence potential is simply proportional to the real part of the superpotential. In addition, even though the shift symmetry is broken by the superpotential, because of the hierarchy between the quintessence energy scale and the Planck scale, such effect does not affect the quintessence dynamics. As an application, we considered the scenario where SUSY is broken at the TeV scale and there is a hidden SQCD axion that plays the role of quintessence. In such scenario, the dark energy scale is given by the electroweak scale and Planck scale, and the cosmic coincidence problem can be ameliorated.

We also proposed another way to circumvent this issue, namely by sequestering the SUSY breaking and quintessence sectors. This approach is based on the picture of a higher dimensional theory where two sectors live on different 3-branes and only communicate with each other through gravity. We showed that indeed quintessence does not acquire a gravitino mass at least at the tree level. Once higher order terms kick in this generally no longer hold and one needs to assume some mechanism in quantum gravity preserves the form of the \Kahler potential  as Eq.(\ref{gQQ}). However, for models with small field displacement, these higher order terms are Planck suppressed and hence do not disturb the dynamics of the quintessence field.

The second main issue one needs to consider is the observational constraints from various gravitational tests like the fifth force constraint. In particular, the strongest source of the coupling between quintessence and matter stems from the exponential factor in the fermionic mass term Eq.(\ref{fermionmass}). In most models, this gives a strong constraint on the quintessence field range, and as we showed in Eq.(\ref{wDeviation}), how much the equation of state parameter can deviate from -1 is constrained by the quintessence field displacement. Therefore, in order to build quintessence models that can be observationally distinguishable from pure cosmological constant, it seems that one needs to ensure quintessence field does not appear in the exponential factor. In the case with shift symmetry, because the \Kahler potential Eq.(\ref{ShiftKahler}) does not depend on the imaginary part of the quintessence superfield which plays the role of the slow-roll quintessence field, the quintessence field does not appear in the exponential factor and hence the observational constraint on matter-quintessence coupling, $\alpha$, defined in Eq.(\ref{alpha}) can be satisfied even for large field displacement. In the sequestered scenario, because the quintessence field still appear in the exponential factor, field displacement is strictly limited by the fifth force constraint and it would be a challenging task to observationally distinguish such models from pure cosmological constant through equation of state parameter.

\section*{Acknowledgement}
We would like to thank Raymond Co, Hajime Fukuda, Keisuke Harigaya, Shigeki Matsumoto, Robert McGehee, Yasunori Nomura, and Tsutomu Yanagida for comments and discussions. HM was supported by the U.S. DOE under Contract DE-AC02-05CH11231, and by the NSF under grants PHY1316783 and PHY-1638509. HM was also supported by the JSPS Grant-in-Aid for Scientific Research (C) (No. 26400241 and 17K05409), MEXT Grant-in-Aid for Scientific Research on Innovative Areas (No. 15H05887, 15K21733), and by WPI, MEXT, Japan. CIC was supported by the NSF East Asia and Pacific Summer Institutes program and thanks Kavli Institute for the Physics and Mathematics of the Universe for hospitality, where part of this work was done.

\addcontentsline{toc}{section}{References}
\bibliographystyle{utphys}
\bibliography{Refs}

\end{document}